\documentclass[twocolumn,showpacs,amsmath,amssymb,a4paper,superscriptaddress]{revtex4}

\usepackage{amsmath,amssymb,bbm,epsfig}

\newcommand{\be}{\begin{equation}}
\newcommand{\ee}{\end{equation}}
\newcommand{\ba}{\begin{array}}
\newcommand{\ea}{\end{array}}
\newcommand{\bqa}{\begin{eqnarray}}
\newcommand{\eqa}{\end{eqnarray}}

\newcommand{\ket}[1]{\ensuremath{| #1 \rangle}}
\newcommand{\fig}[1]{Fig.~\ref{#1}}
\newcommand{\eq}[1]{Eq.~(\ref{#1})}

\begin{document}

\title{
Simulability and regularity of complex quantum systems}

\author{Hannah Venzl}
\affiliation{Physikalisches Institut, Albert-Ludwigs-Universit\"at Freiburg, Hermann-Herder-Str. 3, D-79104 Freiburg, Germany}
\author{Andrew J. Daley}
\affiliation{Institut f\"ur Theoretische Physik, Universit\"at Innsbruck, Technikerstr. 25, A-6020 Innsbruck, Austria}
\affiliation{Institute for Quantum Optics and Quantum Information of the Austrian Academy of Sciences, A-6020 Innsbruck, Austria}
\author{Florian Mintert}
\affiliation{Physikalisches Institut, Albert-Ludwigs-Universit\"at Freiburg, Hermann-Herder-Str. 3, D-79104 Freiburg, Germany}
\author{Andreas Buchleitner}
\affiliation{Physikalisches Institut, Albert-Ludwigs-Universit\"at Freiburg, Hermann-Herder-Str. 3, D-79104 Freiburg, Germany}

\begin{abstract}
We show that the 
transition from regular to chaotic spectral statistics 
in interacting many-body quantum 
systems 
has an unambiguous signature 
in the 
distribution of Schmidt coefficients
dynamically generated from a generic initial state, and thus
limits the efficiency of
the t-DMRG algorithm.
\end{abstract}


\maketitle
Complexity is paradigmatic 
in many areas not alone in physics, but equally so in the life and 
social sciences and in economics \cite{albeverio}. The characteristic property of a ``complex system" resides in the difficulty of its efficient 
simulation through reduction to manageable size. More formally, {\it e.g.} the minimum length of an 
algorithm designed to simulate the system under study 
can serve as a quantitative measure of complexity \cite{kolmogorov}. Complexity 
thus has its very tangible counterpart in the numerical overhead required for an accurate simulation, and implies a challenge for computational physics: 
any time we succeed to minimize that overhead
we actually prove that the underlying complexity is smaller than anticipated. 

Complex systems abound in nature, from interacting many-particle 
systems to deterministic chaos in few degrees of freedom \cite{lichtenberg}, and from macroscopic to microscopic scales. 
On the quantum level, complexity has its counterpart in complex spectral structures, described by random matrix theory and by now well-established 
in nuclear \cite{giannoni} and atomic physics \cite{madronero:263601}, mesoscopics \cite{PhysRevB.54.10841}, and photonics \cite{stoeckmann,zyss}. Recently, ``designed complexity'' moved into reach for state of the art experiments 
on ultracold atoms in periodic optical potentials, and considerable effort is devoted to implementing solid state Hamiltonians with unprecedented control 
in such systems \cite{jaksch:3108,greiner:415,jaksch:315,lewenstein:56}. Since analytical treatments are often unavailable to describe such many-particle 
dynamics, efficient numerical tools are 
in need, and ``efficient'' means here that the required numerical resources as memory and/or execution time scale favorably as compared to the 
exponential growth of Hilbert space dimension with system size.

For 1D systems, 
the advent of Density Matrix Renormalization Group methods \cite{white:2863,schollwock:259} has aided efficient simulation considerably. 
There the underlying idea
lies in the construction of a suitable local basis that makes it possible to represent a system state in terms of
significantly fewer basis states than the total dimension of Hilbert space suggests.
More specifically, this approach makes use of a truncated Matrix Product State (MPS) ansatz \cite{fannes:443}, to reduce the number of coefficients required 
to specify the state to a manageable number. 
In particular in perturbative regimes, where a system has a natural basis, such techniques work very successfully,
and the ground states of 1D systems with local Hamiltonians are typically well represented in this form \cite{vidal:147902,verstraete:094423}.
Also dynamics can be tackled by time-dependent Density Matrix Renormalization Group techniques (t-DMRG). These methods have been shown to work well   
for low-energy initial states \cite{vidal:147902,daley:p04005,white:076401}, giving rise to near-equilibrium time-evolution. For a generic initial state, 
however, they may work only for
short times \cite{schuch:030504}, and apparently depend strongly on the properties of the Hamiltonian.

Here we investigate the connection between complexity and t-DMRG methods: Can t-DMRG efficiently simulate complex many-particle 
systems in general, or does the efficiency of t-DMRG rather identify parameter regimes where the dynamics -- governed by the underlying spectral structure -- is actually
rather ``regular''? To assess this question, we need independent measures of complexity, which, on the spectral level, are provided by the theory of quantum 
chaos \cite{giannoni}.
As we show in the following, the appearance of ``globally irregular" vulgo ``chaotic"  
spectral structure implies 
the breakdown of simulability by t-DMRG techniques.
More specifically, we observe a characteristic qualitative change of the distribution of dynamically generated Schmidt coefficients  that 
directly reflects the defining property of complexity, namely the effective {\em irreducibility} of the Hilbert space dimension: 
Any basis truncation leads to the rapid accumulation of uncontrollable errors in the simulation.

Specifically, we investigate the tilted Bose-Hubbard model. 
Since our results rely on universal 
statistical properties 
that do
not 
depend on system specificities, the
observed phenomena are expected to hold more generally.

%%%%%%%%%%%%%%%%%%%%%%%%%%%%%%%%%%%%%%%%%%%%%%%%%%%%%%
The tilted Bose-Hubbard model we consider here corresponds to bosonic atoms trapped in the lowest band of an optical lattice subject to an additional static field. For $m$ lattice sites, the system is described by the Hamiltonian
\begin{equation}
H=-\frac{J}{2}\sum_{l=1}^{m} (\hat a_{l+1}^\dagger \hat a_l+h.c.)+\frac{U}{2}\sum_{l=1}^{m} \hat n_l(\hat n_l-1)+F\sum_{l=1}^{m} l \hat n_l\ ,
\label{Ehamiltonian}
\end{equation}
where $\hat a_{l}^\dagger$ ($\hat a_l$)
are creation (annihilation) operators for a particle at lattice site $l$, and
$\hat n_l=\hat a_{l}^\dagger\hat a_l$ is the
corresponding number operator.
The system's dynamical properties are characterized
by the tunneling constant $J$, the interaction strength $U$, and a linear potential with strength $F$,
resulting, {\it e.g.}, from gravity when the optical lattice is tilted with respect to gravitational equipotential lines.
The rich dynamics of this system ranges from
perfect oscillations \cite{gustavsson:080404,fattori:080405} 
with the Bloch period $T_B=2\pi/F$, for weak interactions, to chaotic dynamics for situations in which interaction and tunneling are of comparable order of magnitude \cite{kolovsky:056213}.

As mentioned above, simulability via t-DMRG methods relies on the possibility to effectively decimate the Hilbert space of the system in the course of the propagation, by using a truncated MPS representation.
This is based on the Schmidt decomposition for every possible bipartite splitting of the system \cite{schmidt:433},
\begin{equation}
\ket{\psi}=\sum_{\alpha=1}^{\chi_{AB}}\lambda_\alpha \ket{\Phi_\alpha^{[A]} }\ket{\Phi_\alpha^{[B]}}\ ,
\label{Esd}
\end{equation}
where $A$ and $B$ label the two subsystems with Schmidt eigenstates
$\{\ket{\Phi_\alpha^{[A]}}\}$ and $\{\ket{\Phi_\alpha^{[B]}}\}$, and
the real coefficients $\lambda_\alpha$, with normalization condition $\sum_\alpha\lambda_\alpha^2=1$, describe the quantum correlations between the two subsystems.
In the truncated MPS, we make an approximation to the state by setting an upper bound $\chi$ on $\chi_{AB}$, retaining only those eigenstates with the largest $\lambda_\alpha$.  This approximation is good if the $\lambda_\alpha$ 
decay rapidly as a function of the index $\alpha$, when ordered from largest to smallest.
Thus, it relies on the assumption that the entanglement between two parts of the system, as, {\it{e.g.}}, measured by the von Neumann entropy
\begin{equation}
S=-\sum_{\alpha}\lambda_\alpha^2 \log_2 \lambda_\alpha^2\ ,
\label{vne}
\end{equation}
 is never too large.
The von Neumann entropy thus also provides an estimate for the value of $\chi$ required to represent the state, and its behaviour is often used as an indicator of simulability of the system. If $S$ grows rapidly as a function of time, simulation of the system will be difficult, whereas if $S$ is bounded during the dynamics, then we can fix $\chi$ and compute the dynamics over long time periods. 

We now consider this approximation for the time evolution of various initial states, in different parameter regimes of the Hamiltonian, \eq{Ehamiltonian}, above.
In the following we always discuss that bipartition that yields the
largest von Neumann entropy, since this 
limits the efficiency of the t-DMRG algorithm (However, other bipartitions also yield qualitatively similar results.).
In order to permit our subsequent comparison with the system's spectral properties, 
we choose a relatively small system, beginning with
eight bosons in eight neighboring sites at the centre of a lattice of length 64, with Dirichlet boundary conditions, for different values of $U/J$ and $F/J$. 
Note that for $F/J\gtrsim1$, the particles hardly spread on the time scale of the simulation,
which makes it possible to compare the dynamical behavior of the larger system to the spectrum of the Floquet-Bloch operator on nine sites.
However, for very weak static fields
the atoms rather diffuse through the system, and travel into initially unoccupied regions,
so that the correspondence between spectral statistics and t-DMRG is lost.
Therefore, 
our subsequent analysis does not extend to very small values of $F/J$, where diffusion sets in.

\begin{figure}
\begin{center}
\includegraphics[width=.4\textwidth]
{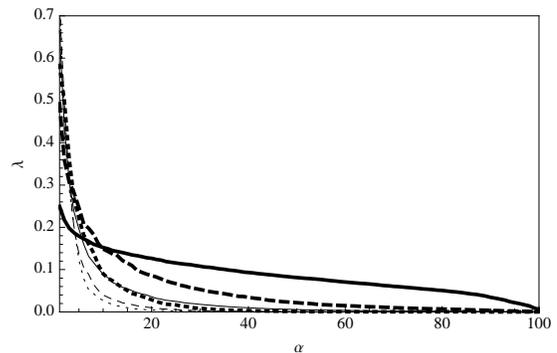}\\
\caption{Distribution of average Schmidt coefficients $\lambda_\alpha$, obtained from averaging over ten initial states of eight particles initially distributed over eight lattice sites. The total grid size was 
$64$, 
to eliminate boundary effects. Shown are the $\chi=100$ largest 
average Schmidt coefficients,
sorted in 
descending order,
for $U/J=1$ (thick lines) and $U/J=10$ (thin lines), and for different values of the static 
tilt $F/J=1.0$ (solid), $1.5$ (dashed), $2.0$ (dotted), 
at $t=4.776\times T_B$. 
}
\label{Fdistrlambda}
\end{center}
\end{figure}

Fig.~\ref{Fdistrlambda} shows the averaged Schmidt coefficients after an evolution time 
$t=4.776\times T_B$, 
with the average 
taken over 
ten separable initial states of the form $\ket{\psi_{\rm init}}\sim \ket{n_1}\otimes\ket{n_2}\otimes\ldots\otimes\ket{n_8}$ 
(each with a different random realisation of occupation numbers $n_\ell$ of eight initially occupied sites).
The distributions can be divided into two qualitatively different categories:
For strong interaction $U/J=10$, few comparably large coefficients dominate the distribution which exhibits a rapidly diminishing tail.
In contrast, for weak interaction $U/J=1$, the distribution shows a slowly decaying tail with many non-negligible coefficients of comparable weight. 

The eligibility of two such distributions for 
the basis truncation
required to apply 
the t-DMRG algorithm is 
fundamentally different: while 
a distribution of the former type 
allows the truncation of the major portion of the Schmidt basis with virtually vanishing loss of accuracy,
dropping a few basis states in the latter case already will lead to a sizable error. 
That is, we can identify situations that can be described efficiently with t-DMRG,
but there are 
parameter regimes of the {\em same} system where a faithful representation of the solution must be spanned 
essentially by the {\em complete} Hilbert space. {\em Any} numerical simulation then is plagued by highly  
unfavorable scaling.
In particular, as immediately evident from Fig.~\ref{Fdistrlambda},
MPS basis truncation at $\chi=100$ 
for $U/J=1$, $F/J=1$ 
enforces a rapid decrease of the Schmidt coefficients for $\alpha>80$, which, in turn, will induce artifacts in the simulated dynamics (as we confirmed by running computations for varying $\chi$). 

Such transition from efficient to inefficient representation in a MPS basis has its cause in a sudden and pronounced 
transition in the underlying spectral structure, as we will now evidence by direct inspection of 
the spectrum of the time evolution (or Floquet) operator 
\begin{equation}
U(T_B)=\hat T \text{exp}\left(-i\int_0^{T_B}\tilde H(t)dt\right)
\label{Efb}
\end{equation}
generated by 
the time-dependent, transformed Hamiltonian
\begin{equation}
H(t)=-\frac{J}{2}\sum_{l=1}^{m} (e^{iFt}\hat a_{l+1}^\dagger \hat a_l+h.c.)+\frac{U}{2}\sum_{l=1}^{m} \hat n_l(\hat n_l-1)\ ,
\label{Ehamiltoniant}
\end{equation}
($\hat T$ denotes time ordering).
Due to the translational invariance of $H(t)$ with periodic boundary conditions,  
$U(T_B)$
decomposes into the direct sum of operators labeled by distinct values of quasimomentum $\kappa$ \cite{kolovsky:056213},
and so does its spectrum.
The statistical analysis therefore requires a diagonalization at 
fixed quasimomentum,
and we chose $\kappa=0$ here.

\begin{figure}
\begin{center}
\hspace{-.4cm}\includegraphics[width=.425\textwidth]
{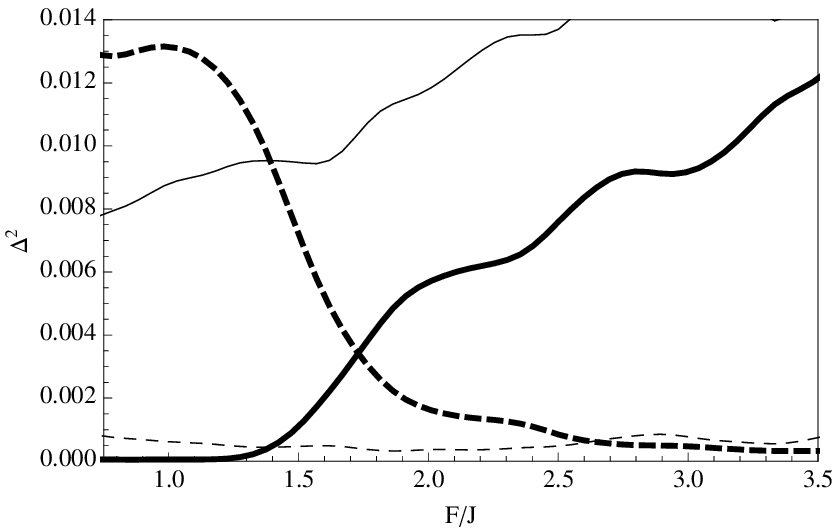}
\caption{
Mean square deviation $\Delta^2$, \eq{Edev}, of the 
Floquet-Bloch operator's (\ref{Efb}) nearest neighbor distribution from Poissonian (dashed lines) and Wigner-Dyson statistics (solid lines), for eight particles on 
nine lattice sites, as a function of $F/J$. $U/J = 1$ (thick lines), $10$ (thin lines).}
\label{Fstatistics}
\vspace{1cm}
\includegraphics[width=.4\textwidth]
{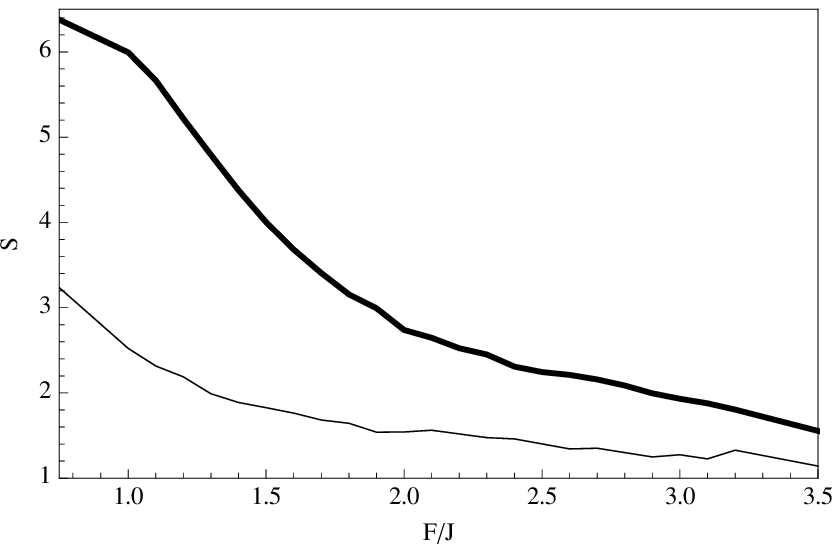}\\
\vspace{-4.6cm}\hspace{3.0cm}\includegraphics[width=0.21\textwidth]
{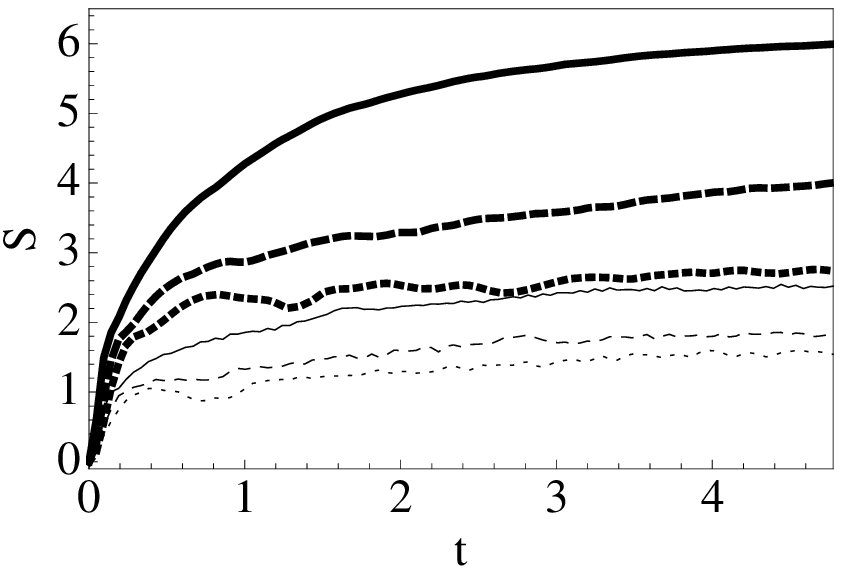}\\\vspace{2.5cm}
\caption{Many-particle entanglement vs. $F/J$, 
measured by the 
average von Neumann entropy $S$, extracted from t-DMRG simulations of eight particles for ten different initial states at
$t=4.776\times T_B$. $U/J = 1$ (thick), $10$ (thin line). Inset: $S$ vs. $t$, for various tilt strengths $F/J=1.0$ (solid), $1.5$ (dashed), $2.0$ (dotted lines),
and the same values of $U/J$.}
\label{Fvne}
\vspace{1cm}
\includegraphics[width=.4\textwidth]
{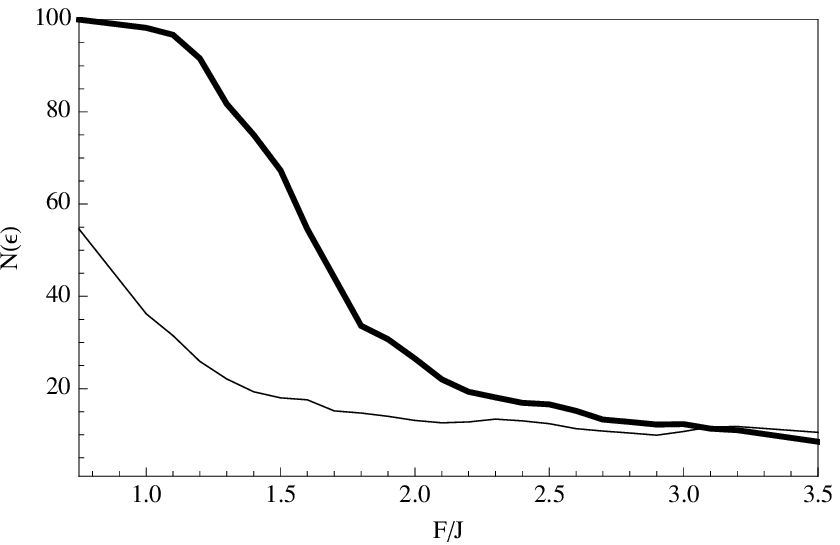}
\caption{Average number of Schmidt coefficients larger than $\epsilon=0.01$,
vs. $F/J$, after a simulation time $t=4.776\times T_B$. $U/J = 1$ (thick), $10$ (thin lines).}
\label{Feps}
\end{center}
\end{figure}

The integrated level spacing distribution $I(s)$ of the eigen-frequencies of $U(T_B)$ 
allows to distinguish regular spectral structure (tantamount of weakly coupled basis states), described 
by 
Poissonian statistics \cite{berry:375}
\begin{equation}
I_P(s)=\int_0^sP(s^\prime)ds^\prime=1-\text{exp}\left({-s}\right)\ ,
\label{Epoisson}
\end{equation}
from a chaotic 
spectrum 
that obeys Wigner-Dyson statistics \cite{giannoni},
\begin{equation}
I_W(s)=1-\text{exp}\left({-\pi s^2/4}\right)\ .
\label{Egoe}
\end{equation}
\fig{Fstatistics} displays the mean square deviation
\begin{equation}
\Delta^2=\frac{1}{\cal{N}}\int ds
\left(f(s)-I(s)\right)^2
\label{Edev}
\end{equation}
of the numerically obtained spectra
with respect to 
$I_W(s)$ 
and $I_P(s)$. 

For $U/J=10$, 
the system obeys Poissonian statistics, irrespective of $F/J$.
However, in the case of $U/J=1$ 
there are three different regimes:
for $F/J\lesssim1.3$, the deviation from (irregular/chaotic) Wigner-Dyson statistics is negligible, 
for $1.3\lesssim F/J\lesssim 2$ the distribution changes its character,
and turns (regular) Poissonian for $F/J\gtrsim 2$.
This transition 
is also reflected in the entropy of Schmidt coefficients 
$S$, \eq{vne},
that is depicted as a function of $F/J$ after a simulation time of $t=4.776\times T_B$ in \fig{Fvne}.
The inset shows the entropy as a function of time, for different values of the static field strength $F/J$.
In the regime of regular level statistics, $S$ grows only initially, whereas it keeps 
increasing 
in the chaotic regime. As a matter of fact,
the saturation of $S$ for $F/J \lesssim 1$ and $U/J=1$ is once 
again a numerical artifact of the truncation at $\chi=100$: 
The simulation only approximates the real values of $S$ 
from below, i.e., at given $t$, $S$ grows when
increasing $\chi$. This behavior of $S$ is perfectly consistent with our observations on Fig.~\ref{Fdistrlambda}. Note that the slight increase
of $S$ towards small values of $F/J$, for $U/J=10$ in Fig.~\ref{Fvne}, corresponds to a narrow distribution of the Schmidt coefficients in 
Fig.~\ref{Fdistrlambda}, and therefore does not hinder efficient simulation -- in perfect agreement with the regular spectral structure in this regime 
spelled out by Fig.~\ref{Fstatistics}.

While the signature of the sharp ``chaos-transition" observed in Fig.~\ref{Fstatistics} is somewhat smoother in the corresponding behaviour of the 
von Neumann entropy in Fig.~\ref{Fvne}, the number of  Schmidt coefficients larger than a certain threshold $\epsilon$ turns out to provide an equally 
sensitive probe as the spectral statistics, as demonstrated in Fig.~\ref{Feps} (for $\epsilon=0.01$):
Whereas in the regular regime
($U/J=10$, or $U/J=1$ with $F/J\gtrsim2$)
less than $20\%$ of the coefficients exceed the threshold $\epsilon$, 
essentially all of them contribute in the chaotic regime ($U/J=1$, $F/J\lesssim1.3$).
That is, whereas the t-DMRG algorithm allows an efficient simulation of the Bose-Hubbard dynamics in the regular regime,
a basis truncation in the chaotic regime will rapidly lead to sizable errors in the simulation. An accurate description requires large numerical efforts that scale exponentially, much as the system size itself.
This observation also holds for larger systems, where t-DMRG is a powerful tool in the regime of regular spectral structure, and
where an exact treatment of the dynamics becomes unfeasible: In t-DMRG calculations with 20 atoms in 20 sites  we observe precisely the
same characteristic changes in the distribution of Schmidt coefficients as observed in Figs.~\ref{Fdistrlambda} and \ref{Feps}. 
Therefore, the {\em inefficiency} of t-DMRG simulations, quantified by the statistical quantities evaluated in Figs.~\ref{Fstatistics} and \ref{Feps} is an
unambiguous indicator of the underlying complexity of the many-particle dynamics. 

Moreover, since the non-existence of a natural basis is a rather generic feature of quantum chaotic systems, we expect
that our observations directly translate to generic many-body quantum systems.
In fact, we conjecture that the distributions of Schmidt coefficients of typical states in spectrally regular and chaotic systems also exhibit universal
features \cite{kubotani:240501} -- much like the energy level distributions 
of regular and chaotic quantum systems.

We thank C. Kollath for helpful conversations.
F. M. gratefully acknowledges financial support by the Alexander von Humboldt foundation.
Work in Innsbruck was supported by Austrian Science foundation through SFB F15 and Project I118\_N16 (EuroQUAM\_DQS).

\bibliography{ref_abu.bib}

\end{document}